\documentclass[aps,pra,showpacs,twocolumn,superscriptaddress,amsmath,amssymb]{revtex4}
\usepackage{graphicx,graphics}
\usepackage{amsmath,sidecap}
\usepackage{color}
\begin{document}

\title{Double-quantum spin vortices in SU(3) spin-orbit coupled Bose gases}
\author{Wei Han}
\affiliation{Key Laboratory of Time and Frequency Primary Standards, National Time Service Center, Chinese Academy of Sciences, Xi'an 710600, China}
\affiliation{Beijing National Laboratory for Condensed Matter Physics, Institute of Physics, Chinese Academy of Sciences, Beijing
100190, China}
\author{Xiao-Fei Zhang}
\affiliation{Key Laboratory of Time and Frequency Primary Standards, National Time Service Center, Chinese Academy of Sciences, Xi'an 710600, China}
\author{Shu-Wei Song}
\affiliation{State Key Laboratory Breeding Base of Dielectrics Engineering, Harbin University of Science and Technology, Harbin 150080, China}
\author{Hiroki Saito}
\affiliation{Department of Engineering Science, University of Electro-Communications, Tokyo 182-8585, Japan}
\author{Wei Zhang\footnote{wzhangl@ruc.edu.cn}}
\affiliation{Department of Physics, Renmin University of China, Beijing 100872, China}
\author{Wu-Ming Liu\footnote{wliu@iphy.ac.cn}}
\affiliation{Beijing National Laboratory for Condensed Matter Physics, Institute of Physics, Chinese Academy of Sciences, Beijing 100190, China}
\author{Shou-Gang Zhang\footnote{szhang@ntsc.ac.cn}}
\affiliation{Key Laboratory of Time and Frequency Primary Standards, National Time Service Center, Chinese Academy of Sciences, Xi'an 710600, China}

%\date{\today}
\begin{abstract}
We show that double-quantum spin vortices, which are characterized by doubly quantized circulating spin currents and unmagnetized filled cores, can exist in the ground states of SU(3) spin-orbit coupled Bose gases. It is found that the SU(3) spin-orbit coupling and spin-exchange interaction play important roles in determining the ground-state phase diagram. In the case of effective ferromagnetic spin interaction, the SU(3) spin-orbit coupling induces a three-fold degeneracy to the magnetized ground state, while in the antiferromagnetic spin interaction case, the SU(3) spin-orbit coupling breaks the ordinary phase rule of spinor Bose gases, and allows the spontaneous emergence of double-quantum spin vortices. This exotic topological defect is in stark contrast to the singly quantized spin vortices observed in existing experiments, and can be readily observed by the current magnetization-sensitive phase-contrast imaging technique.
\end{abstract}

 \maketitle

\section{Introduction}
The recent experimental realization of synthetic spin-orbit (SO) coupling in ultracold quantum gases~\cite{Spielman,Jing-Zhang,Zwierlein,Shuai-Chen,Engels,Jing-Zhang2,Shuai-Chen2,YPChen,Jing-Zhang3,Shuai-Chen3} is considered as an important breakthrough, as it provides new possibilities for ultracold quantum gases to be used as quantum simulation platforms, and paves a new route towards exploring novel states of matter and quantum phenomena \cite{Niu,Galitski,Dalibard,Goldman,Hui-Zhai,Hui-Zhai2,Yirev, Jing-Zhang4,Congjun-Wu}. It has been found that the SO coupling can not only stabilize various topological defects, such as half-quantum vortex, skyrmion, composite soliton and chiral domain wall, contributing to the design and exploration of new functional materials \cite{Congjun-Wu2,Hui-Hu,Malomed,Wei-Han}, but also lead to entirely new quantum phases, such as magnetized phase and stripe phase \cite{Hui-Zhai3,Tin-Lun-Ho,Yun-Li}, providing support for the study of novel quantum dynamical phase transitions \cite{Xiong-Jun-Liu,Polkovnikov} and exotic supersolid phases~\cite{Yun-Li2,Martone,Kuei-Sun}.

All the intriguing features mentioned above are based on the characteristics that the SO coupling (either of the NIST~\cite{Spielman}, Rashba~\cite{Hui-Zhai3} or Weyl~\cite{Spielman2} types) makes the internal states coupled to their momenta via the SU(2) Pauli matrices. However, if the (pseudo)spin degree of freedom involves more than two states, the SU(2) spin matrices cannot describe completely all the couplings among the internal states. For example, a direct transition between the states $|1\rangle$ and $|-1\rangle$ is missing in a three-component system \cite{Lewenstein,Hui-Zhai3}. From this sense, an SU(3) SO coupling with the spin operator spanned by the Gell-Mann matrices is more effective in describing the internal couplings among three-component atoms \cite{Galitski2,Lewenstein}. The SU(3) SO coupled system has no analogue in ordinary condensed matter systems, hence may lead to new quantum phases and topological defects.

In this article, we show that a new type of topological defects, double-quantum spin vortices, can exist in the ground states of SU(3) SO coupled Bose-Einstein condensates (BECs). It is found that the SU(3) SO coupling leads to two distinct ground-state phases, a magnetized phase or a lattice phase, depending on the spin-exchange interaction being ferromagnetic or antiferromagnetic. In the magnetized phase, the SU(3) SO coupling leads to a ground state with three-fold degeneracy, in stark contrast to the SU(2) case where the degeneracy is two, thus may offer new insights into quantum dynamical phase transitions \cite{Xiong-Jun-Liu}. In the lattice phase, the SU(3) SO coupling breaks the ordinary phase requirement $2\mathrm{w}_{0}=\mathrm{w}_{1}+\mathrm{w}_{-1}$ for ordinary spinor BECs, where $\mathrm{w}_{i}$ is the winding number of the $i$-th spin component~\cite{Ueda,Ohmi,Machida}, and induces three types of exotic vortices with cores filled by different magnetizations. The interlaced arrangement of these vortices leads to the spontaneous formation of multiply quantized spin vortices with winding number $2$. This new type of topological defects can be observed in experiments using magnetization-sensitive phase-contrast imaging technique.

\section{SU(3) spin-orbit coupling}
We consider the $F=1$ spinor BECs with SU(3) SO coupling. Using the mean-field approximation, the Hamiltonian can be written in the Gross-Pitaevskii form as
\begin{eqnarray}
\mathcal{H}=\int d\mathbf{r}\left[\mathbf{\Psi }^{\dag}\left(-\frac{\hbar ^{2}\boldsymbol{%
\nabla }^{2}}{2m}+\mathcal{V}_{\text{so}}\right)\mathbf{\Psi }+\frac{c_{0}}{2}n^2+\frac{c_{2}}{2}|\mathbf{F}|^2\right],\label{Hamiltonian}
\end{eqnarray}
where the order parameter $\mathbf{\Psi}=[\Psi_{1}(\mathbf{r}),\Psi_{0}(\mathbf{r}),\Psi_{-1}(\mathbf{r})]^\top$
is normalized with the total particle number $N=\int d\mathbf{r} \mathbf{\Psi}^{\dag}\mathbf{\Psi}$. The particle density is $n=\sum_{m =1,0,-1}\Psi_{m}^{\ast }(\mathbf{r})\Psi _{m}(\mathbf{r})$, and the spin density vector $\mathbf{F}=(F_{x},F_{y},F_{z})$ is defined by $F_{\nu}(\mathbf{r})=\mathbf{\Psi}^{\dag} f_{\nu}\mathbf{\Psi}$ with $\mathbf{f}=(f_{x},f_{y},f_{z})$ being the vector of the spin-$1$ matrices given in the irreducible representation \cite{Ueda,Stamper-Kurn2,DSWang,DSWang2}. The SO coupling term is chosen as $\mathcal{V}_{\text{so}}=\kappa\mathbf{\lambda}\cdot\mathbf{p}$, where $\kappa$ is the spin-orbit coupling strength, $\mathbf{p} = (p_x, p_y)$ represents 2D momentum, and $\mathbf{\lambda}=(\lambda_{x},\lambda_{y})$ is expressed in terms of $\lambda_{x}=\lambda^{(1)}+\lambda^{(4)}+\lambda^{(6)}$ and $\lambda_{y}=\lambda^{(2)}-\lambda^{(5)}+\lambda^{(7)}$, with $\lambda^{(i)} (i=1,...8)$ being the Gell-Mann matrices, i.e., the generators of the SU(3) group~\cite{Arfken}. Note that the SU(3) SO coupling term in the Hamiltonian involves all the pairwise couplings between the three states. This is distinct from the previously discussed SU(2) SO coupling in spinor BECs, where the states $\Psi_{1}(\mathbf{r})$ and $\Psi_{-1}(\mathbf{r})$ are coupled indirectly \cite{Hui-Zhai3,Zhi-Fang-Xu,Zhihao-Lan}. The parameters $c_{0}$ and $c_{2}$ describe the strengths of density-density and spin-exchange interactions, respectively.

\begin{figure}[tbp]
\includegraphics[width=0.92\columnwidth\vspace{0cm} \hspace{0cm}]{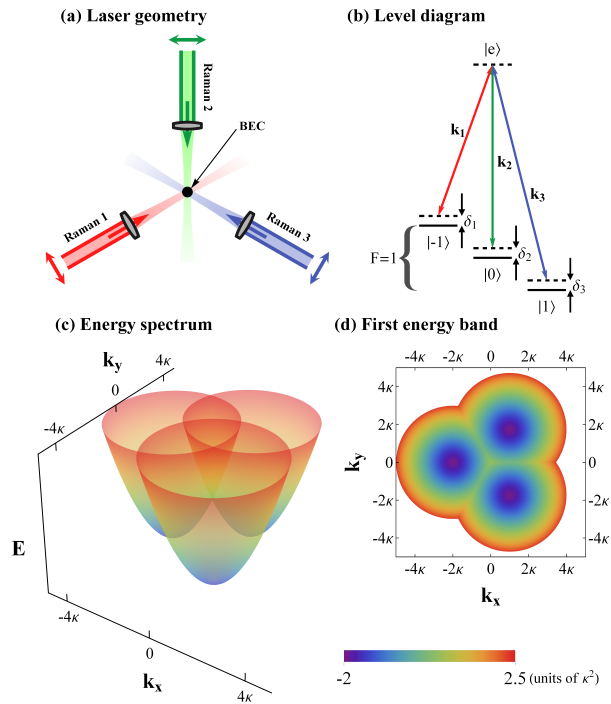}
\caption{(Color online) Scheme for creating SU(3) spin-orbit coupling in spinor BECs. (a) Laser geometry. Three laser beams with different frequencies and polarizations, intersecting at an angle of $2\pi/3$, illuminate the cloud of atoms. (b) Level diagram. Each of the three Raman lasers dresses one hyperfine Zeeman level from $|F=1,m_{F}=1\rangle$, $|F=1,m_{F}=0\rangle$ and $|F=1,m_{F}=-1\rangle$ of the $^{\text{87}}$Rb $5S_{1/2}$, $F=1$ ground
state. $\delta_{1}$, $\delta_{2}$ and $\delta_{3}$ correspond to the detuning in the Raman transitions. (c) Triple-well dispersion relation. The SU(3) spin-orbit coupling induces three discrete minima of the single-particle energy band on the vertices of an equilateral triangle in the $k_{x}$-$k_{y}$ plane. (d) Projection of the first energy band on a 2D plane. Units with $\hbar=m=1$ are used for simplicity.}\label{fig1}
\end{figure}

The Hamiltonian with SU(3) SO coupling can be realized using a similar method of Raman dressing as in the SU(2) case~\cite{Spielman,Jing-Zhang3,Chuangwei-Zhang}. As shown in Fig. \ref{fig1}(a), three laser beams with different polarizations and frequencies, intersecting at an angle of $2\pi/3$, are used for the Raman coupling. Each of the three Raman lasers dresses one hyperfine spin state from the $F=1$ manifold ($|F=1,m_{F}=1\rangle$, $|F=1,m_{F}=0\rangle$ and $|F=1,m_{F}=-1\rangle$) to the excited state $| e \rangle$ [See Fig. \ref{fig1}(b)]. When the standard rotating wave approximation is used and the excited state is adiabatically eliminated due to far detuning, one can obtain the effective Hamiltonian in Eq.~(\ref{Hamiltonian}), as discussed in Appendix \ref{App:Deri}.

By diagonalizing the kinetic energy and SO coupling terms, we can obtain the single-particle energy spectrum, which can provide useful information about the ground state of Bose condensates. For the SU(2) case, it is known that the single-particle spectrum with the NIST type SO coupling acquires either a single or two minima, depending on the strength of the Raman coupling \cite{Spielman}, while for the case of Rashba type there exist an infinite number of minima locating on a continuous ring in momentum space \cite{Galitski3}. For the SU(3) SO coupling discussed here, we find that there are in general three discrete minima residing on the vertices of an equilateral triangle [See Figs. \ref{fig1}(c)-\ref{fig1}(d)]. This unique property of the energy band implies the possibility of a three-fold degenerate many-body magnetized state \cite{Xiong-Jun-Liu} or a topologically nontrivial lattice state, depending on the choices among the three minima made by the many-body interactions.

\section{Phase diagram}
Next, we discuss the phase diagram of the many-body ground states. For the case of SU(2) SO coupling, it is shown that two many-body ground states, magnetized state and stripe state, can be stabilized in a homogeneous system \cite{Yun-Li,Shuai-Chen2,Hui-Zhai3}. Although the Rashba SO coupling provides infinite degenerate minima in the single-particle spectrum, a many-body ground state condensed in one or two points in momentum space is always energetically favorable due to the presence of spin-exchange interaction~\cite{Hui-Zhai3}. As a result, a lattice state with the condensates occupying three or more momentum points for SU(2) SO coupling is unstable, unless a strong harmonic trap is introduced \cite{Santos,Hui-Hu,Zhi-Fang-Xu}.

For the present case of SU(3) SO coupling, we first analytically calculate the possible ground states using a variational approach with a trial wave function $\mathbf{\Psi}=\alpha_{1}\Psi_{1}+\alpha_{2}\Psi_{2}+\alpha_{3}\Psi_{3}$, where

\begin{subequations}
\label{groundstate}
\begin{eqnarray}
\Psi _{1} &=&\frac{1}{\sqrt{3}}\left(
\begin{array}{c}
1 \\
1 \\
1%
\end{array}%
\right) e^{-i2\kappa x},   \label{groundstate1} \\
\Psi _{2} &=&\frac{1}{\sqrt{3}}\left(
\begin{array}{c}
e^{-i\frac{\pi }{3}} \\
e^{i\frac{\pi }{3}} \\
e^{i\pi}%
\end{array}%
\right) e^{i\kappa (x-\sqrt{3}y)},  \label{groundstate2} \\
\Psi _{3} &=&\frac{1}{\sqrt{3}}\left(
\begin{array}{c}
e^{i\frac{\pi }{3}} \\
e^{-i\frac{\pi }{3}} \\
e^{i\pi}%
\end{array}%
\right) e^{i\kappa (x+\sqrt{3}y)},  \label{groundstate3}
\end{eqnarray}
\end{subequations}
correspond to the many-body states with all particles condensing on one of the three minima of the single-particle spectrum, and $\alpha_{i=1,2,3}$ are expansion coefficients. Substituting Eqs.~(\ref{groundstate1})-(\ref{groundstate3}) into the interaction energy functional
\begin{eqnarray}
E=\int d\mathbf{r}\left(\frac{c_{0}}{2}n^2+\frac{c_{2}}{2}|\mathbf{F}|^2\right),\label{interaction energy}
\end{eqnarray} one obtains
\begin{eqnarray}
\frac{E}{N}=\left(\frac{c_{0}}{2}+\frac{4c_{2}}{9}\right)\bar{n}-\frac{7c_{2}}{9\bar{n}}\sum_{i\neq j}|\alpha_{i}|^{2}|\alpha_{j}|^{2},\label{interaction energy2}
\end{eqnarray}
where $\bar{n}=|\alpha_{1}|^{2}+|\alpha_{2}|^{2}+|\alpha_{3}|^{2}$ is the mean particle density. By minimizing the interaction energy with respect to the variation of $|\alpha_{i}|^{2}$, one finds that the spin-exchange interaction plays an important role in determining the phase diagram.

When $c_{2}\!>\!0$, it favors $|\alpha_{1}|^2\!=\!|\alpha_{2}|^2\!=\!|\alpha_{3}|^2\!=\!\bar{n}/3$, indicating that the ground state is a triangular lattice phase with an equally weighted superposition of the three single-particle minima. On the other hand, as $c_{2}\!<\!0$, the system prefers a state with either $|\alpha_{1}|^2\!=\!\bar{n}$, $|\alpha_{2}|^2\!=\!|\alpha_{3}|^2\!=\!0$, or $|\alpha_{2}|^2\!=\!\bar{n}$, $|\alpha_{1}|^2\!=\!|\alpha_{3}|^2\!=\!0$, or $|\alpha_{3}|^2\!=\!\bar{n}$, $|\alpha_{1}|^2\!=\!|\alpha_{2}|^2\!=\!0$, indicating that the ground state occupies one single minimum in the momentum space, and corresponds to a three-fold degenerate magnetized phase.
\begin{figure}[tbp]
\includegraphics[width=0.95\columnwidth\vspace{0cm} \hspace{0cm}]{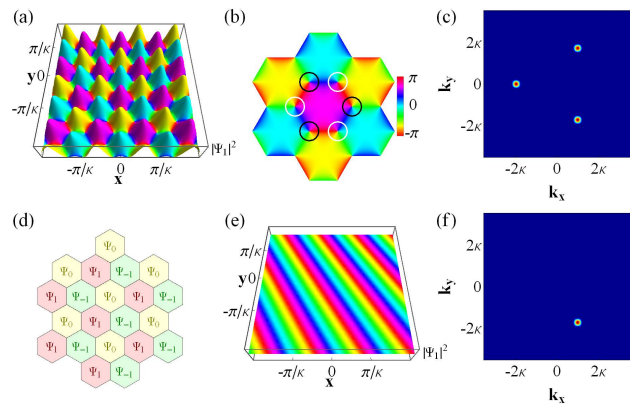}
\caption{(Color online) Two distinct phases present in SU(3) spin-orbit coupled BECs. (a)-(d) The topologically nontrivial lattice phase for antiferromagnetic spin interaction ($c_{2}>0$) with (a) the density and phase of $\Psi_{1}$ represented by heights and colors, (b) the phase within one unit cell showing the positions of vortices (white circles) and antivortices (black circles), (c) the corresponding momentum distributions, and (d) the structural schematic drawing of the phase separation. (e)-(f) The three-fold degenerate magnetized phase for ferromagnetic spin interaction ($c_{2}<0$) with (e) the density and phase distributions of $\Psi_{1}$ and (f) the corresponding momentum distributions.}\label{fig2}
\end{figure}

Note that the variational wave function Eqs. (\ref{groundstate1})-(\ref{groundstate3}) is a good starting point as the SO coupling is strong enough to dominate the chemical potential. For the case with weak SO coupling, one must rely on numerical simulations to determine the many-body ground state. In such a situation, we find a stripe phase with two minima in momentum space occupied for $c_2 \gg \kappa^2$, which will be discussed latter.

The many-body ground states can be numerically obtained by minimizing the energy functional associated with the Hamiltonian Eq.~(\ref{Hamiltonian}) via the imaginary time evolution method. It is found that the numerical results are consistent with the analytical analysis discussed above for rather weak interaction with $c_2 \lesssim \kappa^2$. Figure \ref{fig2} illustrates the two possible ground states of spinor BECs with SU(3) SO coupling. When $c_{2}>0$, the three components are immiscible and arranged as an interlaced triangular lattice with the spatial translational symmetry spontaneously broken [See Figs. \ref{fig2}(a)-\ref{fig2}(d)]. This lattice is topologically nontrivial and embedded by vortices and antivortices as shown in Fig. \ref{fig2}(b). From this result, we conclude that a lattice phase can be stabilized in a uniform SU(3) SO coupled BEC, which is in clear contrast to the SU(2) case where a strong harmonic trap is required~\cite{Santos,Hui-Hu,Zhi-Fang-Xu}. More details on the structure of vortices as well as their unique spin configurations will be investigated later. On the other hand, as $c_{2}<0$, the three components are miscible, and the system forms a magnetized phase with the spatial transitional symmetry preserved but the time-reversal symmetry broken [See Figs. \ref{fig2}(e)-\ref{fig2}(f)]. This magnetized phase occupies one of the three minima of the single-particle spectrum by spontaneous symmetry breaking, hence is three-fold degenerate instead of doubly degenerate in the SU(2) case~\cite{Yun-Li,Xiong-Jun-Liu}.
\begin{figure}
\includegraphics[width=0.4\textwidth]{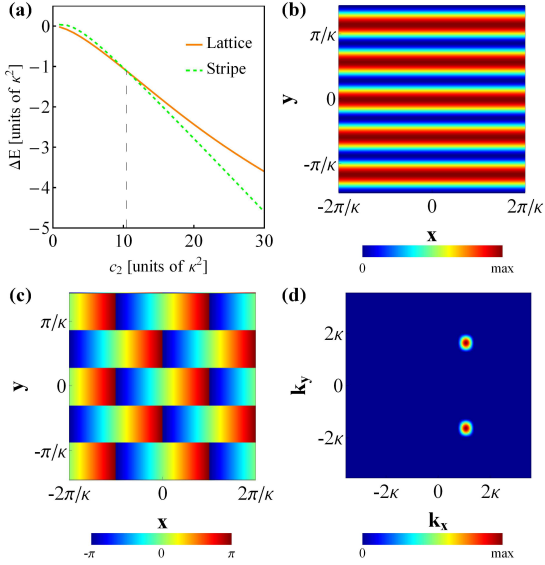}
\caption{(color online) (a) Energy comparison between the lattice and stripe phases. The energy difference $\Delta E$ between the numerical simulation and the variational calculation are shown by solid (lattice state) and dashed (stripe state) lines. (b)-(d) The ground-state density, phase and momentum distributions of the stripe phase with the parameters $c_{2}=20 \kappa^{2}$ and $c_{0}=10 c_{2}$.}
\label{fig3}
\end{figure}

For strong antiferromagnetic spin interaction with $c_2 \gg \kappa^2$, however, a stripe phase is identified with two of three minima occupied in the momentum space. We take the states with two or three minima occupied in the momentum space as trial wave functions, and perform imaginary time evolution to find their respective optimized ground state energy. A typical set of results are summarized in Figure \ref{fig3}(a), showing the energy comparison with different values of interatomic interactions. Obviously, one finds that the stripe phase will has lower energy than the lattice phase when the interatomic interaction exceeds a critical value. Due to the finite momentum in vertical direction of the stripe [See Fig. \ref{fig3}(d)], both the spatial translational and time-reversal symmetries are broken [See Figs. \ref{fig3}(b)-\ref{fig3}(c)]. This is distinct from the stripe phase induced by SU(2) SO coupling, where the time-reversal symmetry is preserved \cite{Hui-Zhai3}.

\section{Phase requirement}
The vortex configuration of spinor BECs depends on the phase relation between the three components. We next discuss the influence of SO coupling on the phase requirement of the vortex configuration. We first assume that the spinor order parameter of a vortex in the polar coordinate $(r, \theta)$ can be described as
\begin{eqnarray}
\psi_{j}(r,\theta)=\phi_{j}e^{i\mathrm{w}_{j}\theta+\alpha_{j}},\label{spinor order parameter}
\end{eqnarray}
where $j=0,\pm1$ and $\phi_{j}\geq0$.

\subsection{Without spin-orbit coupling}
In the absence of SO coupling, the phase-dependent terms in the Hamiltonian are
\begin{eqnarray}
H^{\mathrm{phase}}&=&E_{\mathrm{kin}}^{\mathrm{phase}}+E_{\mathrm{int}}^{\mathrm{phase}}\notag \\
&=&\!\!\!\!-\frac{1}{2}\!\!\int{\!\!\mathbf{\Psi}^{*}\frac{1}{r^2}\frac{\partial^2}{\partial\theta^2}\mathbf{\Psi} d\mathbf{r}}\!+\!2c_{2}\!\!\int{\!\!\Re({\psi_{1}^{*}\psi_{-1}^{*}\psi_{0}^{2}})d\mathbf{r}},\label{phase-dependent terms}
\end{eqnarray}
where the first term results from the kinetic energy and the second from the spin-exchange interaction. Substituting Eq. (\ref{spinor order parameter}) into (\ref{phase-dependent terms}), one obtains
\begin{align}
& E_{\mathrm{kin}}^{\mathrm{phase}}=\sum_{j=1,0,-1}\mathrm{w}_{j}^2\int\frac{\pi\phi_{j}^2}{r}dr, \label{Ekin} \\
& E_{\mathrm{int}}^{\mathrm{phase}}=2c_{2}\int\phi_{1}\phi_{-1}\phi_{0}^2rdr \notag \\
& \int\cos\left[(\mathrm{w}_{1}-2\mathrm{w}_{0}+\mathrm{w}_{-1})\theta
+(\alpha_{1}-2\alpha_{0}+\alpha_{-1})\right]d\theta. \label{Eint}
\end{align}
It is easy to read from Eq. (\ref{Ekin}) that the system favors small winding numbers energetically. Moreover, from Eq.~(\ref{Eint}) the energy minimization requires the winding number and phase satisfy the following relations
\begin{subequations}
\label{phase requirement1}
\begin{eqnarray}
\mathrm{w}_{1}-2\mathrm{w}_{0}+\mathrm{w}_{-1} &=&0, \label{winding number1} \\
\alpha_{1}-2\alpha_{0}+\alpha_{-1} &=&n\pi, \label{phase1}
\end{eqnarray}
\end{subequations}
where $n$ is odd for $c_{2}>0$ and even for $c_{2}<0$. The phase requirement of Eq. (\ref{winding number1}) indicates that the following types of winding combination, such as $\langle \pm1,\times,0\rangle$, $\langle 0,\times,\pm1\rangle$, $\langle \pm1,0,\mp1\rangle$, $\langle \pm1,\pm1,\pm1\rangle$, $\langle \pm2,\pm1,0\rangle$ and $\langle 0,\pm1,\pm2\rangle$ are allowed in a spinor BEC, where the symbol ``$\times$" denotes the absence of the $\Psi_{0}$ component.

\subsection{With SU(2) spin-orbit coupling}
For the case of SU(2) SO coupling, we take the Rashba type as an example, and write the Hamiltonian as
\begin{eqnarray}
E_{\mathrm{soc}}\!=\!\!\!\int\!\!\!\kappa \mathbf{\psi }^{\dag}\!\! \left(\!\!
\begin{array}{ccc}
0 & -i\partial _{x}-\partial _{y}  & 0 \\
-i\partial _{x}+\partial _{y}  & 0 & -i\partial
_{x}-\partial _{y}  \\
0 & -i\partial _{x}+\partial _{y}  & 0%
\end{array}%
\!\!\right)\!\!\mathbf{\psi }d\mathbf{r}, \label{Rashba Hamiltonian}
\end{eqnarray}
where $\mathbf{\psi}=[\psi_{1},\psi_{0},\psi_{-1}]^\top$. Substituting Eq. (\ref{spinor order parameter}) into (\ref{Rashba Hamiltonian}), one can obtain
\begin{align}
& E_{\mathrm{soc}}=\!\!\!\!\int\!\!\! dr d\theta \Big[(\phi_{0}r\partial_{r}\phi_{1}-\mathrm{w}_{1}\phi_{0}\phi_{1}) e^{i[(\mathrm{w}_{1}-\mathrm{w}_{0}+1)\theta+(\alpha_{1}-\alpha_{0}-\frac{\pi}{2})]}  \notag \\
&-(\phi_{1}r\partial_{r}\phi_{0}+\mathrm{w}_{0}\phi_{1}\phi_{0}) e^{-i[(\mathrm{w}_{1}-\mathrm{w}_{0}+1)\theta+(\alpha_{1}-\alpha_{0}-\frac{\pi}{2})]}  \notag \\
&+(\phi_{0}r\partial_{r}\phi_{-1}+\mathrm{w}_{-1}\phi_{0}\phi_{-1}) e^{i[(\mathrm{w}_{-1}-\mathrm{w}_{0}-1)\theta+(\alpha_{-1}-\alpha_{0}-\frac{\pi}{2})]}  \notag \\
&-(\phi_{-1}r\partial_{r}\phi_{0}-\mathrm{w}_{0}\phi_{-1}\phi_{0}) e^{-i[(\mathrm{w}_{-1}-\mathrm{w}_{0}-1)\theta+(\alpha_{-1}-\alpha_{0}-\frac{\pi}{2})]}\Big].
\end{align}
In order to minimize the SO coupling energy, it is preferred that
\begin{subequations}
\label{phase requirement2}
\begin{eqnarray}
\mathrm{w}_{1}-\mathrm{w}_{0}+1 &=&0, \label{winding number2a} \\
\mathrm{w}_{-1}-\mathrm{w}_{0}-1 &=&0, \label{winding number2b} \\
\alpha_{1}-\alpha_{0}-\frac{\pi}{2} &=&m\pi, \label{phase2a} \\
\alpha_{-1}-\alpha_{0}-\frac{\pi}{2} &=&n\pi. \label{phase2b}
\end{eqnarray}
\end{subequations}
Then the SO coupling energy is rewritten as
\begin{align}
& E_{\mathrm{soc}}=2\pi\!\!\int\![\phi_{0}r\partial_{r}\phi_{1}-\phi_{1}r\partial_{r}\phi_{0}
-(\mathrm{w}_{1}+\mathrm{w}_{0})\phi_{0}\phi_{1}]dr\cos{m\pi}  \notag \\
&+2\pi\!\!\int\![\phi_{0}r\partial_{r}\phi_{-1}-\phi_{-1}r\partial_{r}\phi_{0}
+(\mathrm{w}_{-1}+\mathrm{w}_{0})\phi_{0}\phi_{-1}]dr\cos{n\pi},\label{Esoc-su2}
\end{align}
where $m$ and $n$ are odd or even, which can be determined by minimizing the energy expressed in Eq. (\ref{Esoc-su2}). It is found that the SU(2) SO coupling does not violate the ordinary requirement on the winding combination in Eq. (\ref{winding number1}), but introduces further requirements in Eqs.~(\ref{winding number2a})-(\ref{winding number2b}). As a result, while $\langle -1,0,1\rangle$, $\langle -2,-1,0\rangle$ and $\langle 0,1,2\rangle$ are still allowed, some winding combinations such as $\langle \pm1,\pm1,\pm1\rangle$, $\langle \pm1,\times,0\rangle$, $\langle 0,\times,\pm1\rangle$, $\langle 1,0,-1\rangle$, $\langle 2,1,0\rangle$ and $\langle 0,-1,-2\rangle$ are forbidden. Obviously, one can see that the SO coupling break the chiral symmetry, thus may lead to chiral spin textures.

\subsection{With SU(3) spin-orbit coupling}
For the case of SU(3) SO coupling, the effective Hamiltonian can be written as
\begin{eqnarray}
E_{\mathrm{soc}}\!=\!\!\!\int\!\!\!\kappa \mathbf{\psi }^{\dag} \!\!\left(\!\!
\begin{array}{ccc}
0 & -i\partial _{x}-\partial _{y}  & -i\partial _{x}+\partial _{y} \\
-i\partial _{x}+\partial _{y}  & 0 & -i\partial
_{x}-\partial _{y}  \\
-i\partial _{x}-\partial _{y} & -i\partial _{x}+\partial _{y}  & 0%
\end{array}%
\!\!\right)\!\!\mathbf{\psi }d\mathbf{r}. \label{SU3 SOC Hamiltonian}
\end{eqnarray}
Substituting Eq. (\ref{spinor order parameter}) into (\ref{SU3 SOC Hamiltonian}), we get
\begin{align}
& E_{\mathrm{soc}}=\!\!\int\!\!\!dr d\theta \Big[(\phi_{0}r\partial_{r}\phi_{1}-\mathrm{w}_{1}\phi_{0}\phi_{1}) e^{i[(\mathrm{w}_{1}-\mathrm{w}_{0}+1)\theta+(\alpha_{1}-\alpha_{0}-\frac{\pi}{2})]}  \notag \\
&-(\phi_{1}r\partial_{r}\phi_{0}+\mathrm{w}_{0}\phi_{1}\phi_{0}) e^{-i[(\mathrm{w}_{1}-\mathrm{w}_{0}+1)\theta+(\alpha_{1}-\alpha_{0}-\frac{\pi}{2})]}  \notag \\
&+(\phi_{0}r\partial_{r}\phi_{-1}+\mathrm{w}_{-1}\phi_{0}\phi_{-1}) e^{i[(\mathrm{w}_{-1}-\mathrm{w}_{0}-1)\theta+(\alpha_{-1}-\alpha_{0}-\frac{\pi}{2})]}  \notag \\
&-(\phi_{-1}r\partial_{r}\phi_{0}-\mathrm{w}_{0}\phi_{-1}\phi_{0}) e^{-i[(\mathrm{w}_{-1}-\mathrm{w}_{0}-1)\theta+(\alpha_{-1}-\alpha_{0}-\frac{\pi}{2})]}  \notag \\
&+(\phi_{-1}r\partial_{r}\phi_{1}+\mathrm{w}_{1}\phi_{-1}\phi_{1}) e^{i[(\mathrm{w}_{1}-\mathrm{w}_{-1}-1)\theta+(\alpha_{1}-\alpha_{-1}-\frac{\pi}{2})]}  \notag \\
&-(\phi_{1}r\partial_{r}\phi_{-1}-\mathrm{w}_{-1}\phi_{1}\phi_{-1}) e^{-i[(\mathrm{w}_{1}-\mathrm{w}_{-1}-1)\theta+(\alpha_{1}-\alpha_{-1}-\frac{\pi}{2})]}\Big].
\end{align}
By minimizing the SO coupling energy, one obtains the following relations
\begin{subequations}
\label{phase requirement3}
\begin{eqnarray}
\mathrm{w}_{1}-\mathrm{w}_{0}+1 &=&0, \label{winding number3a} \\
\mathrm{w}_{-1}-\mathrm{w}_{0}-1 &=&0, \label{winding number3b} \\
\mathrm{w}_{1}-\mathrm{w}_{-1}-1 &=&0, \label{winding number3c} \\
\alpha_{1}-\alpha_{0}-\frac{\pi}{2} &=&m\pi, \label{phasephase3a} \\
\alpha_{-1}-\alpha_{0}-\frac{\pi}{2} &=&n\pi, \label{phasephase3b} \\
\alpha_{1}-\alpha_{-1}-\frac{\pi}{2} &=&l\pi. \label{phasephase3c}
\end{eqnarray}
\end{subequations}
Then the SO coupling energy can be rewritten as
\begin{align}
& E_{\mathrm{soc}}=\!2\pi\!\!\int\![\phi_{0}r\partial_{r}\phi_{1}-\phi_{1}r\partial_{r}\phi_{0}
-(\mathrm{w}_{1}+\mathrm{w}_{0})\phi_{0}\phi_{1}]dr\cos{m\pi}  \notag \\
&+\!2\pi\!\!\int\![\phi_{0}r\partial_{r}\phi_{-1}-\phi_{-1}r\partial_{r}\phi_{0}
+(\mathrm{w}_{-1}+\mathrm{w}_{0})\phi_{0}\phi_{-1}]dr\cos{n\pi}  \notag \\
&+\!2\pi\!\!\int\![\phi_{-1}r\partial_{r}\phi_{1}-\phi_{1}r\partial_{r}\phi_{-1}
+(\mathrm{w}_{1}+\mathrm{w}_{-1})\phi_{-1}\phi_{1}]dr\cos{l\pi},\label{Esoc-su3}
\end{align}
where $m$, $n$ and $l$ are odd or even, which can be determined from Eq. (\ref{Esoc-su3}). However, the three winding requirements Eqs.~(\ref{winding number3a})-(\ref{winding number3c}) can not be satisfied simultaneously. Thus the SU(3) SO coupling may choose two out of the three winding requirements for the following three cases:\newline

Case \uppercase\expandafter{\romannumeral1}:
\begin{subequations}
\label{hanhan2}
\begin{eqnarray}
\mathrm{w}_{1}-\mathrm{w}_{0}+1 &=&0, \label{winding number3a1} \\
\mathrm{w}_{-1}-\mathrm{w}_{0}-1 &=&0, \label{winding number3b1} \\
\alpha_{1}-\alpha_{0}-\frac{\pi}{2} &=&m\pi, \label{phasephase3a1} \\
\alpha_{-1}-\alpha_{0}-\frac{\pi}{2} &=&n\pi. \label{phasephase3b1}
\end{eqnarray}
\end{subequations}

Case \uppercase\expandafter{\romannumeral2}:
\begin{subequations}
\label{hanhan2}
\begin{eqnarray}
\mathrm{w}_{1}-\mathrm{w}_{0}+1 &=&0, \label{winding number3a2} \\
\mathrm{w}_{1}-\mathrm{w}_{-1}-1 &=&0, \label{winding number3c2} \\
\alpha_{1}-\alpha_{0}-\frac{\pi}{2} &=&m\pi, \label{phasephase3a2} \\
\alpha_{1}-\alpha_{-1}-\frac{\pi}{2} &=&l\pi. \label{phasephase3c2}
\end{eqnarray}
\end{subequations}

Case \uppercase\expandafter{\romannumeral3}:
\begin{subequations}
\label{hanhan2}
\begin{eqnarray}
\mathrm{w}_{-1}-\mathrm{w}_{0}-1 &=&0, \label{winding number3b2} \\
\mathrm{w}_{1}-\mathrm{w}_{-1}-1 &=&0, \label{winding number3c2} \\
\alpha_{-1}-\alpha_{0}-\frac{\pi}{2} &=&n\pi, \label{phasephase3b2} \\
\alpha_{1}-\alpha_{-1}-\frac{\pi}{2} &=&l\pi. \label{phasephase3c2}
\end{eqnarray}
\end{subequations}
For case \uppercase\expandafter{\romannumeral1}, the winding combination $\langle -1,0,1\rangle$ is allowed, while $\langle 1,0,-1\rangle$ is not allowed, indicating the chiral symmetry is broken. For case \uppercase\expandafter{\romannumeral2} and case \uppercase\expandafter{\romannumeral3}, one can find that the SU(3) SO coupling breaks the ordinary requirement on the winding combination in Eq. (\ref{winding number1}), thus new winding combinations, such as $\langle 0,1,-1\rangle$ and $\langle 1,-1,0\rangle$, are possible.

\section{Vortex configurations}
The vortex configurations of spinor BECs can be classified according to the combination of winding numbers and the magnetization of vortex core \cite{Ohmi,Machida,Ueda}. For example, a Mermin-Ho vortex has winding combination $\langle\pm2,\pm1,0\rangle$ with a ferromagnetic core, where the plus and minus signs represent different chirality of the vortices \cite{Saito}, and the expression of $\langle \mathrm{w}_1, \mathrm{w}_0, \mathrm{w}_{-1} \rangle$ indicates that the components of $\Psi_1$, $\Psi_0$ and $\Psi_{-1}$ in the wave function acquire winding numbers of $\mathrm{w}_1$, $\mathrm{w}_0$ and $\mathrm{w}_{-1}$, respectively. Using this notation, a polar-core vortex has winding combination $\langle\pm1,0,\mp1\rangle$ with an antiferromagnetic core, and a half-quantum vortex has winding combination $\langle\pm1,\times,0\rangle$ with a ferromagnetic core, where the symbol ``$\times$" denotes the absence of the $\Psi_{0}$ component.

\begin{figure}[tbp]
\includegraphics[width=0.8\columnwidth\vspace{0cm} \hspace{0cm}]{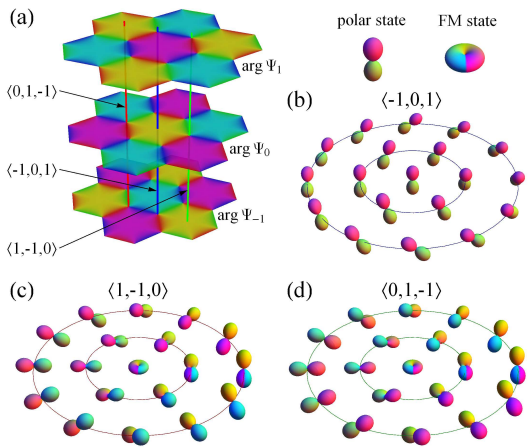}
\caption{(Color online) Vortex configurations in antiferromagnetic spinor BECs with SU(3) spin-orbit coupling. (a) Vortex arrangement among the three components of the condensates. One can identify three types of vortices, including a polar-core vortex with winding combination $\langle -1,0,1\rangle$ (blue line) and two ferromagnetic-core vortices with winding combinations $\langle1,-1,0\rangle$ (green line) and $\langle0,1,-1\rangle$ (red line). (b)-(d) Spherical-harmonic representation of the three types of vortices. The surface plots of $|\mathbf{\Phi}(\theta,\phi)|^{2}$ for (b) the polar-core vortex $\langle -1,0,1\rangle$, (c) the ferromagnetic-core vortex $\langle1,-1,0\rangle$, and (d) the ferromagnetic-core vortex $\langle0,1,-1\rangle$ are shown with the colors representing the phase of $\mathbf{\Phi}(\theta,\phi)$. Here, $\mathbf{\Phi}(\theta,\phi)=\sum_{m =-1}^{1}Y_{1m}(\theta,\phi)\Psi_{m}$ and $Y_{1m}$ is the rank-1 spherical-harmonic function.}\label{fig4}
\end{figure}

In the lattice phase induced by the SU(3) SO coupling with antiferromagnetic spin interaction, there exists three types of vortices: one is a polar-core vortex with winding combination $\langle -1,0,1\rangle$, and the other two are ferromagnetic-core vortices with winding combinations $\langle1,-1,0\rangle$ and $\langle0,1,-1\rangle$ [See Fig. \ref{fig4}(a)]. However, the vortex configurations with opposite chirality of each type, such as $\langle 1,0,-1\rangle$, $\langle-1,1,0\rangle$ and $\langle0,-1,1\rangle$, are not allowed, because the chiral symmetry is intrincically broken in SU(3) SO coupled systems, as discussed in Sec.~\uppercase\expandafter{\romannumeral4}.

Surprisingly, one finds that the two types of ferromagnetic-core vortices $\langle 1,-1,0\rangle$ and $\langle0,1,-1\rangle$ violate the conventional phase requirement $2\mathrm{w}_{0}=\mathrm{w}_{1}+\mathrm{w}_{-1}$ for ordinary spinor BECs~\cite{Ueda,Ohmi,Machida}. This can be understood by noting that the relative phase among different wave function components are no longer uniquely determined by the spin-exchange interaction but also affected by the SU(3) SO coupling, as qualitatively explained in Sec.~\uppercase\expandafter{\romannumeral4}. Thus, the interlaced arrangement of the three types of vortices forms a new class of vortex lattice which has no analogue in systems without SO coupling.

The configurations of the three types of vortices induced by the SU(3) SO coupling with antiferromagnetic interaction are essentially different from those usually observed in ferromagnetic spinor BECs, as can be illustrated by the spherical-harmonic representation~\cite{Ueda}. From Figs.~\ref{fig4}(b)-\ref{fig4}(d) one can find that for the polar-core vortex, the antiferromagnetic order parameter varies continuously everywhere, while for the ferromagnetic-core vortex, the magnetic order parameter acquires a singularity at the vortex core. In contrast, in the ordinary ferromagnetic spinor BECs, the ferromagnetic order parameter varies continuously everywhere for the ferromagnetic-core vortex, but has a singularity at the core for the polar-core vortex \cite{Ueda}.

\section{Double-quantum spin vortices}
Spin vortex is a complex topological defect resulting from symmetry breaking, and is characterized by zero net mass current and quantized spin current around an unmagnetized core \cite{Stamper-Kurn2,Ueda,Saito2,Kawaguchi,Ruostekoski,Gou}. It is not only different from the magnetic vortex found in magnetic thin films \cite{Hubert,Ono,Wiesendanger}, but also from the 2D skyrmion \cite{Bigelow,Shin} due to the existence of singularity in the spin textures \cite{Su-Yi}. Single-quantum spin vortex with the spin current showing one quantum of circulation has been experimentally observed in ferromagnetic spinor BECs~\cite{Stamper-Kurn}. Multi-quantum spin vortices with $l$ ($l\geq2$) quanta circulating spin current, however, are considered to be topologically unstable and have not been discovered yet~\cite{Ueda}.

\begin{figure}[tbp]
\includegraphics[width=0.8\columnwidth\vspace{0cm} \hspace{0cm},clip]{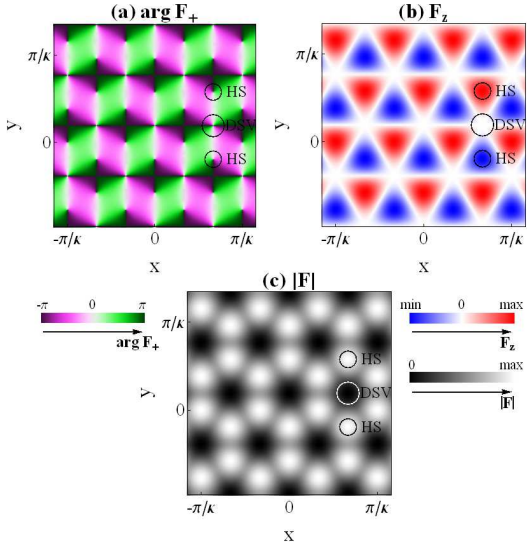}
\caption{(Color online) Double-quantum spin vortex in antiferromagnetic spinor BECs with SU(3) spin-orbit coupling. (a) Spatial maps of the transverse magnetization with colors indicating the magnetization orientation. (b) Longitudinal magnetization. (c) Amplitude of the total magnetization $|\textbf{F}|$. Two kinds of topological defects, double-quantum spin vortex (DSV) and half skyrmion (HS) are marked by big and small circles, respectively. The transverse magnetization orientation arg$F_{+}$ along a closed path (indicated by big circles) surrounding the unmagnetized core shows a net winding of $4\pi$, revealing the presence of a double-quantum spin vortex.}\label{fig5}
\end{figure}

A particularly important finding of our present work is that the polar-core vortex in the lattice phase has a spin current with two quanta of circulation around the unmagnetized core, hence can be identified as a double-quantum spin vortex. Figure~\ref{fig5} presents the transverse magnetization $F_{+}=F_{x}+iF_{y}$, longitudinal magnetization $F_{z}$, and amplitude of the total magnetization $|\textbf{F}|$ in the lattice phase, which are experimentally observable by magnetization-sensitive phase-contrast imaging technique~\cite{expl1}. From these results, one can find two distinct types of topological defects, double-quantum spin vortex (DSV) and half skyrmion (HS) \cite{Brown,Pu}, which correspond to the polar-core vortex with winding combinations $\langle-1,0,1\rangle$ and the ferromagnetic-core vortex with winding combinations $\langle1,-1,0\rangle$ or $\langle0,1,-1\rangle$, respectively. In particular, for the double-quantum spin vortex, the core is unmagnetized and the orientation of the magnetization along a closed path surrounding the core acquires a rotation of $4\pi$. This finding indicates that a regular lattice of multi-quantum spin vortices can emerge spontaneously in antiferromagnetic spinor BECs with SU(3) SO coupling. By exploring the effect of a small but finite temperature, we confirm that the double-quantum spin vortices are robust against thermal fluctuations and hence are observable in experiments, as discussed in Appendix \ref{App:Stab}.

The emergence of spin current with two quanta of circulation can be analytically understood by expanding the wave function obtained by the variational methods around the center of a double-quantum spin vortex. We suppose that the wave function of the lattice phase is written as
\begin{equation}
\psi\!=\!\frac{1}{3}\!\!\left(\!\!
\begin{array}{c}
1 \\
1 \\
1%
\end{array}%
\!\!\right)\!\! e^{-i2\kappa x}+\frac{1}{3}\!\!\left(\!\!
\begin{array}{c}
e^{-i\frac{\pi }{3}} \\
e^{i\frac{\pi }{3}} \\
e^{i\pi }%
\end{array}%
\!\!\right)\!\! e^{i\kappa (x-\sqrt{3}y)}+\frac{1}{3}\!\!\left(\!\!
\begin{array}{c}
e^{i\frac{\pi }{3}} \\
e^{-i\frac{\pi }{3}} \\
e^{i\pi }%
\end{array}%
\!\!\right)\!\! e^{i\kappa (x+\sqrt{3}y)}.  \label{wave function}
\end{equation}
Then one can expand $\psi$ around the center of a vortex with winding number $\langle -1,0,1\rangle$, e.g., at the location of $(x,y)=(0,\pi/(3\sqrt{3}\kappa))$. Substituting $x=\epsilon \cos\theta$ and $y=\pi/(3\sqrt{3}\kappa)+\epsilon \sin\theta$ into $\psi$ and expanding with respect to the infinitesimal $\epsilon$, we obtain
\begin{equation}
\psi =\left(
\begin{array}{c}
-i\kappa e^{-i\theta }\epsilon -\frac{1}{2}\kappa ^{2}e^{i2\theta }\epsilon
^{2} \\
1-\kappa ^{2}\epsilon ^{2} \\
-i\kappa e^{i\theta }\epsilon -\frac{1}{2}\kappa ^{2}e^{-i2\theta }\epsilon
^{2}%
\end{array}%
\right) +O\left( \epsilon ^{3}\right).   \label{EWF}
\end{equation}

\begin{figure}
\includegraphics[width=0.4\textwidth,clip]{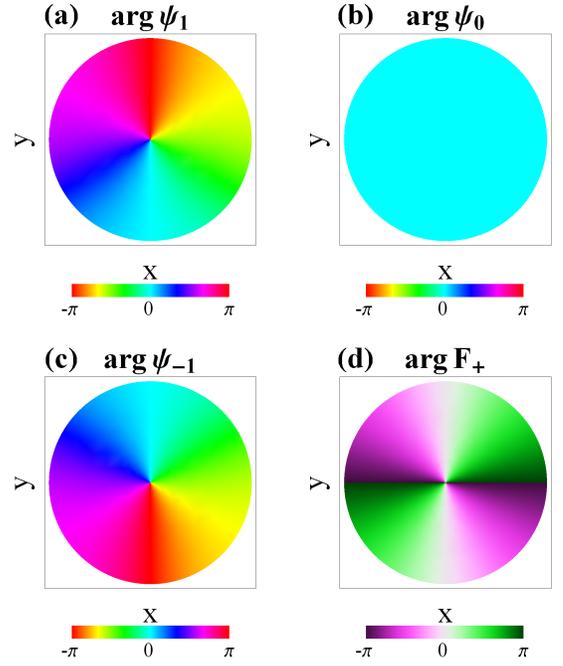}
\caption{(color online) (a)-(c) Phases of the polar-core vortex described by the wave function in Eq. (\ref{EWF}), displaying the winding combination $\langle-1,0,1\rangle$. (d) Direction of the transverse magnetization, indicating the emergence of spin current with two quanta of circulation.}
\label{fig6}
\end{figure}

Notice that the second-order terms with $e^{\pm i2\theta}$ have no essential influence on the phases, thus the winding number for each component can still be represented as $\langle-1,0,1\rangle$ [See Figs. \ref{fig6}(a)-\ref{fig6}(c)]. However, since the first-order terms are canceled out when calculating the transverse magnetization $F_{+}=\sqrt{2}\left[\psi_{1}^{*}\psi_{0}+\psi_{0}^{*}\psi_{-1}\right]$, the second-order terms play a dominant role, leading to the emergence of spin current with two quanta of circulation around an unmagnetized core
\begin{equation}
F_{+}\propto \epsilon^{2}e^{-i2\theta},  \label{Fxy}
\end{equation}
as illustrated in Fig. \ref{fig6}(d).

\section{Conclusion}
To summarize, we have mapped out the ground-state phase diagram of SU(3) spin-orbit coupled Bose-Einstein condensates. Several novel phases  are discovered including a three-fold degenerate magnetized phase, a vortex lattice phase, as well as a stripe phase with time-reversal symmetry broken. We also investigate the influence of SU(3) spin-orbit coupling on the phase requirement of the vortex configuration, and demonstrate that the SU(3) spin-orbit coupling breaks the ordinary phase rule of spinor Bose-Einstein condensates, and allows the spontaneous emergence of stable double-quantum spin vortices. As a new member in the family of topological defects, double quantum spin vortex has never been discovered in any other systems. Our work deepen the understanding of spin-orbit phenomena, and will attract extensive interest of scientists in the cold atom community.

\section*{ACKNOWLEDGMENTS}
This work was supported by NKRDP under grants Nos. 2016YFA0301500, 2012CB821305; NSFC under grants Nos. 61227902, 61378017, 11274009, 11434011, 11434015, 11447178, 11522436, 11547126; NKBRSFC under grant No. 2012CB821305; SKLQOQOD under grant No. KF201403; SPRPCAS under grant Nos. XDB01020300, XDB21030300; JSPS KAKENHI grant No. 26400414 and MEXT KAKENHI grant No. 25103007. W. H. and X.-F. Z. contributed equally to this work.

\appendix
\renewcommand \appendixname{APPENDIX}

\section{DERIVING THE EFFECTIVE HAMILTONIAN}\label{App:Deri}
We consider spinor Bose-Einstein condensates (BECs) illuminated by three Raman laser beams, which couple two of the three hyperfine spin components respectively, as illustrated in Figs. 1(a)-1(b) of the main text. The internal dynamics of a single particle under this scheme can be described by the Hamiltonian
\begin{eqnarray}
H\!\!&=&\!\!\sum_{j=1}^{3}\left(\frac{\hbar^2\mathbf{k}^2}{2m}+\varepsilon_{j}\right)|j\rangle\langle j|+\sum_{l=1}^{n}E_{l}|l\rangle\langle l|  \notag \\
&&\!\!+\sum_{j=1}^{3}\sum_{l=1}^{n}\left[\Omega_{j}e^{i(\mathbf{K}_{j}\cdot\mathbf{r}+\omega_{j}t)}M_{lj}|l\rangle\langle j\rangle+h.c.\right],\label{Single-Particle Hamiltonian}
\end{eqnarray}
where $\hbar\mathbf{k}$ is the momentum of the particles, and $\varepsilon_{j}$ and $E_{l}$ are the energies of the ground and excited states, respectively. In the atom-light coupling term, $\mathbf{K}_{j}$ and $\omega_{j}$ are the wave vectors and frequencies of the three Raman lasers with $\Omega_{j}$ the corresponding Rabi frequencies, and $M_{lj}$ is the matrix element of the dipole transition. One can see that this Hamiltonian is similar to that used in the scheme for creating 2D spin-orbit (SO) coupling in ultracold Fermi gases \cite{Jing-Zhang3}, thus can be readily realized in Bose gases. Taking the standard rotating wave approximation to get rid of the time dependence of the Hamiltonian, and adiabatically eliminating the excited states for far detuning, the Hamiltonian can be rewritten as
\begin{eqnarray}
H\!\!=\!\!\!\left(\!\!
\begin{array}{ccc}\!
\frac{\hbar ^{2}\left( \mathbf{k}+\mathbf{K}_{1}\right) ^{2}}{2m}\!+\!\delta_{1} & \Omega_{12}
& \Omega_{13}  \\
\Omega_{21}  & \frac{\hbar ^{2}\left( \mathbf{k}+\mathbf{K}_{2}\right) ^{2}}{2m}\!+\!\delta_{2}
& \Omega_{23}  \\
\Omega_{31}  & \Omega_{32}  & \frac{\hbar ^{2}\left( \mathbf{k}+\mathbf{K}_{3}\right) ^{2}}{2m}\!+\!\delta_{3}
\end{array}
\!\!\right)\!\!,
\end{eqnarray}
where $\delta_{1}$, $\delta_{2}$ and $\delta_{3}$ are the two-photon detunings, and the real parameters $\Omega_{jj'}=\Omega_{j'j}$ describe the Raman coupling strength between hyperfine ground states $|j\rangle$ and $|j'\rangle$, which can be expressed as \cite{Jing-Zhang3,Thomas}
\begin{eqnarray}
\Omega_{jj'} &=&-\frac{\sqrt{I_{j}I_{j'}}}{\hbar^2c\epsilon_{0}}\sum_{m'}\frac{\langle j'|er_{q}|m'\rangle\langle m'|er_{q}|j\rangle}{\Delta}.\label{Raman coupling strength}
\end{eqnarray}
Here, $I_{j}$ is the intensity of each Raman laser, and $\Delta$ denotes the one-photon detuning. Other parameters $c$, $\epsilon_{0}$ and $e$ in Eq. (\ref{Raman coupling strength}) are the speed of light, permittivity of vacuum and elementary charge, respectively. In Eq.~(\ref{Raman coupling strength}), $q=x,y,z$ is an index labeling the components of $r$ in the spherical basis, and $|m'\rangle$ describes the middle excited hyperfine spin state in the Raman process. For simplicity, we assume $\Omega=\Omega_{12}=\Omega_{13}=\Omega_{23}$, which can always be satisfied by adjusting the system parameters, such as the laser intensity.

Introducing a unitary transformation
\begin{eqnarray}
U=\frac{1}{\sqrt{3}}\left(
\begin{array}{ccc}
1 & 1
& 1  \\
-e^{-i\frac{\pi}{3}}  & -e^{i\frac{\pi}{3}}
& 1  \\
-e^{i\frac{\pi}{3}}  & -e^{-i\frac{\pi}{3}}  & 1
\end{array}
\right)
\end{eqnarray}
and a time-dependent unitary transformation $U(t)=e^{i\left(\frac{\hbar ^{2}\mathbf{K}_{0}^{2}}{2m}+\delta_{2}-\Omega\right)t}$, the effective Hamiltonian becomes
\begin{eqnarray}
H\!=\!\!\left(\!\!
\begin{array}{ccc}
\frac{\hbar ^{2}\mathbf{k}^{2}}{2m}\!+\!\delta_{1}\!-\!\delta_{2} & 0
& 0  \\
0  & \frac{\hbar ^{2}\mathbf{k}^{2}}{2m}
& 0  \\
0  & 0  & \frac{\hbar ^{2}\mathbf{k}^{2}}{2m}\!+\!\delta_{3}\!-\!\delta_{2}\!+\!3\Omega
\end{array}
\!\!\right)\!\!+\!\!\mathcal{V}_{\text{so}},
\end{eqnarray}
where the laser vectors $K_{1}=-K_{0}\mathbf{\hat{e}}_{y}$, $K_{2}=\frac{\sqrt{3}K_{0}}{2}\mathbf{\hat{e}}_{x}+\frac{K_{0}}{2}\mathbf{\hat{e}}_{y}$ and $K_{3}=-\frac{\sqrt{3}K_{0}}{2}\mathbf{\hat{e}}_{x}+\frac{K_{0}}{2}\mathbf{\hat{e}}_{y}$ are defined with $K_{0}=2m\kappa/\hbar$. The spin-dependent uniform potential induced by the Raman detuning $\delta_{i}$ and Raman coupling strength $\Omega$ can be eliminated by applying a Zeeman field, leading to
\begin{eqnarray}
H=\left(
\begin{array}{ccc}
\frac{\hbar ^{2}\mathbf{k}^{2}}{2m}+\epsilon_{1} & 0
& 0  \\
0  & \frac{\hbar ^{2}\mathbf{k}^{2}}{2m}
& 0  \\
0  & 0  & \frac{\hbar ^{2}\mathbf{k}^{2}}{2m}+\epsilon_{2}
\end{array}
\right)+\mathcal{V}_{\text{so}},
\end{eqnarray}
where $\epsilon_{1}=\delta_{1}-\delta_{2}+\Delta_{1}+\Delta_{2}$ and $\epsilon_{2}=\delta_{3}-\delta_{2}-\Delta_{1}+\Delta_{2}+3\Omega$ with $\Delta_{1}$ and $\Delta_{2}$ denoting the linear and quadratic Zeeman energy respectively. By tuning the detuning, the Zeeman energy and the Raman coupling strength, one can reach the regime $\Delta_{1}=\frac{\delta_{3}-\delta_{1}+3\Omega}{2}$ and $\Delta_{2}=\delta_{2}-\frac{\delta_{1}+\delta_{3}+3\Omega}{2}$ which satisfying $\epsilon_{1}=\epsilon_{2}=0$. Then we have
\begin{eqnarray}
H=\frac{\hbar ^{2}\mathbf{k}^{2}}{2m}+\mathcal{V}_{\text{so}},
\end{eqnarray}
which is the single-particle Hamiltonian with SU(3) SO coupling considered in the main text.

\section{STABILITY OF THE DOUBLE-QUANTUM SPIN VORTEX STATES}\label{App:Stab}
\begin{SCfigure*}
\centering
\includegraphics[width=0.38\textwidth]{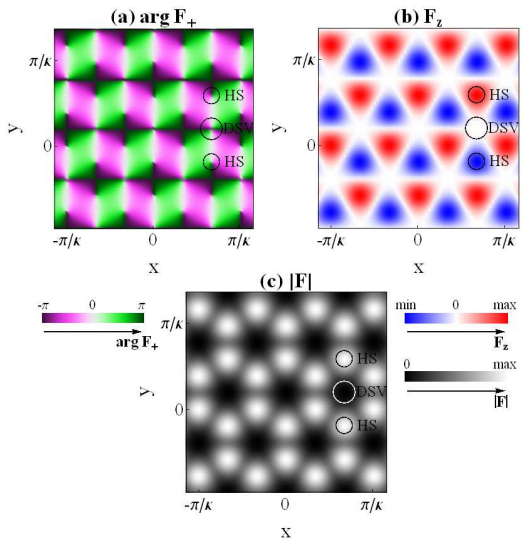}
\caption{(color online) Stable double-quantum spin vortex under a random fluctuation. The images are taken at $t=20 ms$ in the real-time evolution, with thermal fluctuation in the energy scale of $k_{B} T$ with $T\sim 300$ nK. (a) Spatial maps of the transverse magnetization with colors indicating the magnetization orientation. (b) Longitudinal magnetization. (c) Amplitude of the total magnetization $|\textbf{F}|$. It is shown that the double-quantum spin vortices are topologically stable under external fluctuations with a fairly long lifetime of tens of ms.}\label{fig7}
\end{SCfigure*}

In order to verify the stability of the phases discovered in this manuscript, we have explored the effects of a small but finite temperature, and concluded that the double-quantum spin vortex states are robust against the thermal fluctuations. In particular, we considered a random fluctuation $\Delta\phi$ in the real-time evolution of the Gross-Pitaevskii equation, which causes an energy fluctuation about $\Delta E=0.03 E_{g}$ with $E_{g}$ the ground-state energy. An estimation shows that this level of fluctuation corresponds to the energy scale $k_{B} T$ with $T\sim 300$ nK, which is higher enough for a usual system of Bose-Einstein condensates in realistic experiments. According to numerical simulations, we find that the structure of the double-quantum spin vortex state is stable under this level of fluctuation in tens of millisenconds [See Fig. \ref{fig7}], suggesting that this phase is indeed observable in experiments.\newline\newline


\begin{thebibliography}{99}
\bibitem{expl1} Other experimental techniques, such as selective absorption and time-of-flight measurement, can also be used to observe the double-quantum spin vortex states. We suggest the magnetization-sensitive phase-contrast imaging method just because it has advantage in observing the phase information of the spin-vortex structure.
\end{thebibliography}
\end{document}